\documentclass[%
twocolumn,superscriptaddress,
 amsmath,amssymb,
 aps,
prl,
]{revtex4-1}
\usepackage{graphicx}
\usepackage{dcolumn}
\begin{document}


\title{Many-Body Effects on Nuclear Short Range Correlations}


\author{Haoyu Shang}
\affiliation{
State Key Laboratory of Nuclear Physics and Technology, School of Physics,
Peking University, Beijing 100871, China
}

\author{Jiawei Chen}
\affiliation{
State Key Laboratory of Nuclear Physics and Technology, School of Physics,
Peking University, Beijing 100871, China
}
\author{Rongzhe Hu}
\affiliation{
State Key Laboratory of Nuclear Physics and Technology, School of Physics,
Peking University, Beijing 100871, China
}
\author{Xin Zhen}
\affiliation{
State Key Laboratory of Nuclear Physics and Technology, School of Physics,
Peking University, Beijing 100871, China
}
\author{Chongji Jiang}
\affiliation{
State Key Laboratory of Nuclear Physics and Technology, School of Physics,
Peking University, Beijing 100871, China
}
\author{J.C. Pei}\email{peij@pku.edu.cn}
\affiliation{
State Key Laboratory of Nuclear Physics and Technology, School of Physics,
Peking University, Beijing 100871, China
}
\affiliation{
Southern Center for Nuclear-Science Theory (SCNT), Institute of Modern Physics, Chinese Academy of Sciences, Huizhou 516000,  China
}



\begin{abstract}

We reveal nuclear many-body effects on short range correlations
by {\it ab initio} no-core shell model calculations of the scaling factor $a_2$.
The factor $a_2$ characterizes the abundance of SRC pairs and is linearly related to the
EMC effect.
Our study employs the fifth-order N$^4$LO chiral nuclear force without softening,
enabling to distinguish the influences of nuclear states  with different quantum numbers on SRC.
It is striking to find that $a_2$ is reduced and close in triplet isobaric analog states of neighboring nuclei,
indicating that it is insufficient to estimate SRC abundances by considering only  mean-field shell structures.
This is explained as specific nuclear states suppress the formation of deuteron-like component,
impacting our understandings of the link between high-energy partonic properties and
low-energy nuclear physics.
\end{abstract}


\maketitle

\emph{Introduction.}---
Nuclear short-range correlations (SRC) exhibit an unique link
across energy scales from low-energy to high-energy nuclear physics~\cite{link}.
It was revealed that high momentum nucleons in nuclei tend to form SRC pairs,
with large relative momentum and smaller centre-of-mass (c.m.) momentum~\cite{hen2017nucleon,tropiano2021short}.
The SRC pairs
embody the realistic nucleon-nucleon interaction at short distances and have universal behaviors at high momentum in all nuclei~\cite{frankfurt1993evidence,degli2015medium,weiss2015generalized}.
On the other hand, it is of fundamental interests to understand the quark-gluon effects inside
nuclei by analyzing SRC contributions to nuclear parton distribution functions~\cite{link,emc,hen2017nucleon,clas2019modified}.

SRCs are usually characterized by the scaling factor $a_2(\text{A}/d)$,
which measures the probability
of finding a two-nucleon SRC pair in the nucleus relative to deuteron by inclusive high-energy electron scattering experiments ~\cite{egiyan2006measurement,frankfurt1993evidence,fomin2012new,clas2019modified,li2022revealing}.
It can be calculated by evaluating the ratio of the high-momentum tail of
the nucleon momentum distribution~\cite{ryckebusch2019isospin,weiss2018nuclear,hen2017nucleon}.
The scaling factor $a_2$ is linearly related to the magnitude of EMC effect~\cite{emc}.
Further exclusive measurements find that $np$-SRCs dominate whereas $pp$-SRCs have almost negligible contribution~\cite{shneor2007investigation,korover2014probing,duer2019direct,schmidt2020probing}.
Currently, the probability of finding $np$-SRC pairs in a nucleus composed of $Z$ protons and $N$ neutrons
is evaluated as $L\frac{NZ}{A}$, where $L$ being the Levinger constant\cite{levinger1951high,weiss2015nuclear,weiss2015generalized}.
In principle, the factor $a_2$ is expected to be dependent on specific nuclear structures,
and this issue becomes essential to understand more precise measurements in the future.

The repulsive feature of $nn$ and $pp$ correlation functions at short ranges already appear~\cite{corrfun}, considering
the influence of Pauli principle
in the Fermi gas model or mean-field models, while $np$ pairs are allowed at short ranges.
To extract many-body effects on SRC beyond the mean field, {\it ab initio} calculations of the scaling factor $a_2$ with realistic
nuclear forces are needed.
Actually  many-body wave functions in different nuclei
can be factorized into a product of a universal short-range pair
and a residual $A$-2 part~\cite{cruz2021many}.
To probe short-range structures, \textit{ab initio} calculations require the use of hard nuclear interactions,
which restrict the applicability of many numerical methods.
Variational Monte Carlo calculations (VMC) are able to handle hard interactions and give access to properties of a range
of light nuclei~\cite{wiringa2014nucleon,piarulli2023densities,carlson2015quantum}.
However, VMC is restricted to local interactions~\cite{piarulli2016local,wiringa1995accurate} and currently provide momentum distributions only for the ground state.
Most chiral nucleon-nucleon ($NN$) interactions are developed in momentum space and are non-local~\cite{entem2003accurate,entem2017high,ekstrom2013optimized,epelbaum2005two,reinert2018semilocal}.
In this respect, no-core shell model (NCSM) is an ideal full-configuration many-body method that can handle both local and non-local nuclear forces~\cite{barrett2013ab}.
Furthermore, NCSM is a natural choice to distinguish many-body effects on SRC in nuclear states associated with different spin, parity and isospin.

In this Letter, to study many-body effects on SRC, we develop
NCSM calculations of  $a_2$ in different states of $^6 \text{Li}$.
Specifically, we studied its ground state 1$^+$, as well as the first and second excited states 3$^+$ and 0$^+$.
This is an appealing example to demonstrate many-body effects on SRC, as $^6 \text{Li}$ has a core of $^4$He
plus a valence $np$ pair.
 It is expected that the valence $np$ pair in different configurations
 should have distinct contributions to SRC, in contrast to the tightly bound $^4$He core.
Moreover, the 0$^+$ state in $^6 \text{Li}$, corresponding to ground states in $^6 \text{He}$ and $^6 \text{Be}$,
  are isobaric analog states.
In this Letter, it  is striking to find that the isobaric analog states in three nuclei
have close $a_2$ values although their proton and neutron numbers are very different.
In our NCSM calculations, the latest fifth-order ($\text{N}^4\text{LO}$) nonlocal chiral interaction~\cite{entem2017high}
are employed.
To calculate $a_2$, it is crucial to obtain translationally invariant (ti) one-nucleon momentum distributions~\cite{navratil2004translationally},
and this is realized by unfolding the c.m. motion from the one-nucleon density
in the momentum space.




\emph{Translationally invariant density.}---
NCSM calculations conventionally adopt single-particle coordinates rather than relative coordinates.
As a result, wave functions $\psi(\vec{k}_i)$ include c.m. motion and this is a key issue for calculating the scaling factor $a_2$.
The harmonic oscillator (HO) basis and full $N_{\rm{max}}$ truncation result in
the exact factorization of the wave functions into a c.m. wave function and a ti wave function~\cite{gloeckner1974spurious}:
\begin{equation}
  \psi^{\omega}\left(\vec{k}_{1},\vec{k}_{2},\ldots,\vec{k}_{A}\right)=
  \phi^{\omega}_{{\rm cm}}(\vec{K})
  \psi_{\mathrm{ti}}\left(\vec{k}_{{\rm rel}1},\vec{k}_{{\rm rel}2},\ldots,\vec{k}_{{\rm rel}A}\right),
\end{equation}
where $\vec{K}=\sum_{i=1}^{A} \vec{k}_i$ and $\vec{k}_{\rm rel}=\vec{k}-\frac{\vec{K}}{A}$.
The c.m. wave function is $\phi^{\omega}_{{\rm cm}}(\vec{K})=e^{-K^{2}\nu^{2}/2}(\frac{\nu}{\sqrt{\pi}})^{3/2}$,
where $\nu=\frac{b}{\sqrt{A}}$ with $b=\sqrt{\frac{\hbar}{m\omega}}$ being the HO length.

The space-fixed (sf) density~\cite{giraud2008density,PhysRevC.86.034325} is obtained by
the integral of $|\psi^{\omega}|^2$, which depends on the  HO basis frequency $\omega$.
The ti density $\rho_{{\rm ti}}(\vec{k})$ is obtained by
integral of $|\psi_{\mathrm{ti}}|^2$, which describes the probability of finding a nucleon with momentum $\vec{k}$
 relative to the c.m. of the entire nucleus and does not depend on $\omega$.
These two densities are related by:
\begin{equation}
  \rho_{\rm sf}^\omega(\vec{k})=\int\rho_{{\rm ti}}(\vec{k}-\vec{K}/A)\rho_{\rm cm}^\omega(\vec{K})d^3K,
\end{equation}
where $\rho_{\rm cm}^\omega={\phi^\omega_{\rm cm}}^2(\vec{K})$, which smears out $\rho_{\rm ti}$.

The ti density reflects the intrinsic properties of the nucleus and can be used to extract the scaling factor,
which is obtained by unfolding the c.m. motion from the directly calculated sf density using standard Fourier methods:
\begin{equation}
  \rho_{\mathrm{ti}}(\vec{k})=F\left[\frac{F^{-1}\left[\rho_{\rm{sf}}^{\omega}(\vec{q})\right](\vec{r})}
  {(2\pi)^{3}A^{3}F^{-1}\left[\rho_{\mathrm{cm}}^{\omega}(A\vec{K})\right](\vec{r})}\right](\vec{k}). \label{eq:unfold}
\end{equation}

In calculations of ti densities, there could be oscillations at high momentum by using the Fourier transformations.
To solve this issue, the sf densities from NCSM are first extrapolated with
a smoothing exponent and transformed into  $\tilde{\rho}_{\mathrm{sf}(r)}$. Then oscillations are sufficiently suppressed for heavier nuclei such as $^6$Li.
Finally the angle-averaged ti densities can be derived as,
\begin{equation}
  \rho_{{\rm ti}}(k)=4\pi\int j_{K}(kr)\frac{\tilde{\rho}_{\mathrm{sf}}(r)}{e^{-\frac{r^{2}}{4A^{2}\nu^{2}}}}r^{2}dr. \label{eq:unfoldcal}
\end{equation}

\emph{Extraction of scaling factor $a_2$.}---
Theoretical predictions for the scaling factors are proposed to be obtained by evaluating
ratios of two-nucleon distributions in the limit of infinitely large relative momentum~\cite{chen2017short,lynn2020ab,ryckebusch2019isospin}:
\begin{equation}
  a_{2}(A/d) =\lim_{k\rightarrow\infty}\frac{2}{A}\frac{\rho_{2}(A,k)}{\rho_{2}(d,k)}, \label{eq:scalingfactor}
\end{equation}
where the two-nucleon distribution is defined as $\rho_2(A,k)=\langle\psi|\sum_{i<j}^A \delta(k-k_{ij})|\psi\rangle /(4\pi k^2)$
and is normalized to the number of nucleon pairs.
In momentum space, for $k\rightarrow\infty$ there is a simple relation between the one-nucleon
and the two-nucleon momentum distributions~\cite{weiss2015generalized}:
\begin{equation}
  \begin{aligned}
    \rho_{\mathrm{ti},A}^p(k) &=2\rho_{2}^{pp}(A,k)+\rho_{2}^{np}(A,k)\\
    \rho_{\mathrm{ti},A}^n(k) &=2\rho_{2}^{nn}(A,k)+\rho_{2}^{np}(A,k). \label{eq:onetotwo}
  \end{aligned}
\end{equation}

Consequently, the two-nucleon momentum distributions in Eq.~\eqref{eq:scalingfactor} can be substituted
by one-nucleon momentum distribution to evaluate the scaling factor:
\begin{equation}
  a_{2}(A/d) = \lim_{k\rightarrow\infty}\frac{\rho_{\mathrm{ti},A}(k)}{A\left|\tilde{\psi}_{d}(k)\right|^{2}}
  \approx \frac{\int_{k_0}^{\infty}\rho_{\mathrm{ti},A}(k)k^2 dk}{A\int_{k_0}^\infty\left|\tilde{\psi}_{d}(k)\right|^{2}k^2 dk}.\label{eq:sca_fac_eva}
\end{equation}
Here $k_0$ is typically chosen to be in the same range as fermi momentum ($k_F \approx 1.4$ $\mathrm{fm}^{-1}$).
The integral of ti density is usually adopted for smooth estimations of $a_2$~\cite{weiss2018nuclear,tropiano2021short,ryckebusch2019isospin}.

\begin{figure}[t]
  \includegraphics[width=\columnwidth]{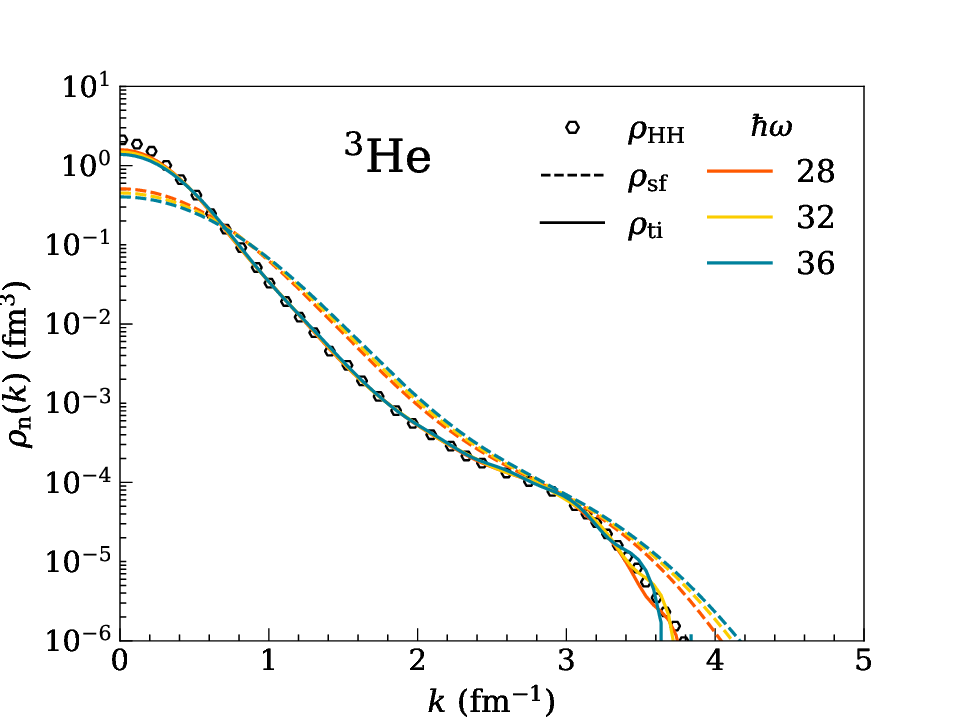}
  \caption{\label{fig:he3benchmark}
   Neutron momentum distribution of $^3\mathrm{He}$ calculated by NCSM using the N$^4$LO nonlocal $NN$ interaction.
  The open-circles represent results obtained by the HH method~\cite{marcucci2019momentum}.
  The dashed and solid lines correspond to the sf and ti densities, respectively,  calculated by NCSM with $N_{\mathrm{max}}$ = 16, and $\hbar \omega$ ranging
  from 24 MeV to 36 MeV.
  }
\end{figure}

\emph{Results and analysis.---}
Firstly we calculate the neutron density distribution of $^3\mathrm{He}$ using NCSM
with the bare N$^4$LO nonlocal NN interaction EMN500~\cite{entem2017high},
to benchmark with the hyperspherical harmonics (HH) method~\cite{marcucci2019momentum}, as shown in Fig.\ref{fig:he3benchmark}.
The NCSM code is taken from Ref.~\cite{michel2021gamow}.
The sf densities are calculated at $N_{\mathrm{max}}$= 16 with $\hbar \omega$ ranging from 24 MeV to 36 MeV.
The chosen value of $N_{\mathrm{max}}$ is sufficient to get convergence in densities.
The HH method, by employing Jacobi coordinates, avoids the c.m. problem~\cite{marcucci2020hyperspherical}.
Then ti densities from NCSM calculations are benchmarked with the HH results~\cite{marcucci2019momentum} to verify our procedure for unfolding the c.m. motion.
Note that ti densities obtained using Eq.~\eqref{eq:unfoldcal} exhibit small oscillations when the momentum $k$ is larger than 3.8 fm$^{-1}$, while
this issue doesn't exist in $^6$Li.
Besides, the $\hbar\omega$ dependence of the sf densities is removed in the ti densities after unfolding the c.m. motion.
It can be seen that the ti densities agree quite well with the HH results within a reasonable range of momentum.

\begin{figure}[t]
  \includegraphics[width=\columnwidth]{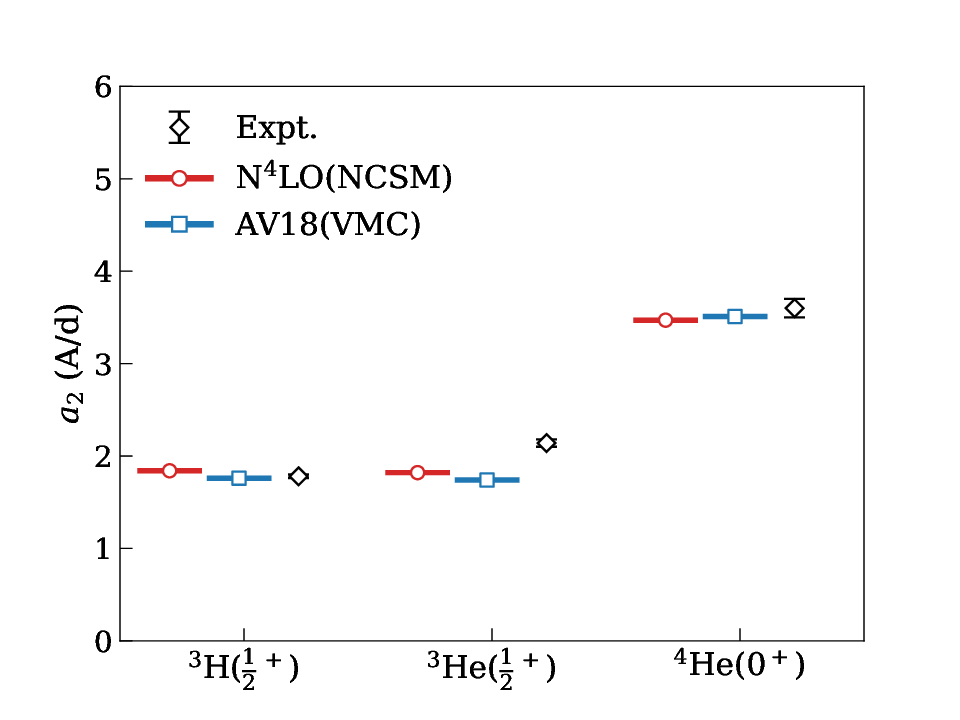}
  \caption{\label{fig:scaling_He3}
  The estimated scaling factor $a_2$ in ground states of $^3$H, $^3$He and $^4$He by NCSM calculations with the N$^4$LO chiral nuclear force,
  to compare with experiments~\cite{clas2019modified,li2022revealing}.
Results using densities from VMC calculations with the AV18 nuclear force are also shown.
  }
\end{figure}

In Fig.~\ref{fig:scaling_He3},  the resulted scaling factor $a_2$ of $^3$H, $^3$He and $^4$He are shown, as given by Eq.\eqref{eq:sca_fac_eva}.
For comparison, experimental values~\cite{clas2019modified,li2022revealing} and results with the local phenomenological force AV18~\cite{wiringa1995accurate}  are also displayed.
The AV18 results are obtained by using densities from VMC calculations~\cite{wiringa2014nucleon}, which
includes three-body forces. In our NCSM calculations, the nonlocal chiral force N$^4$LO~\cite{entem2017high} is adopted.
Note that over a reasonable range of $k_0$, the ratio of the integrated density
remains relatively unchanged.  We choose $k_0=2$ $\mathrm{fm}^{-1}$ to evaluate the scaling factor in the following, as suggested in Refs.~\cite{tropiano2021short,ryckebusch2019isospin}.
It is shown that the scaling factors $a_2$ appear to be not sensitive to different nuclear interactions.
As pointed out in Ref.~\cite{tropiano2021short}, the high momentum dependence cancels out in the ratios.
The calculated  $a_2$ values  agree well with experiments although the results of  $^3$He are slightly underestimated,
due to additional contributions of $pp$-SRC in the measurement of $^3$He rather than the isospin symmetry breaking~\cite{li2022revealing}.

\begin{figure}[t]
  \includegraphics[width=\columnwidth]{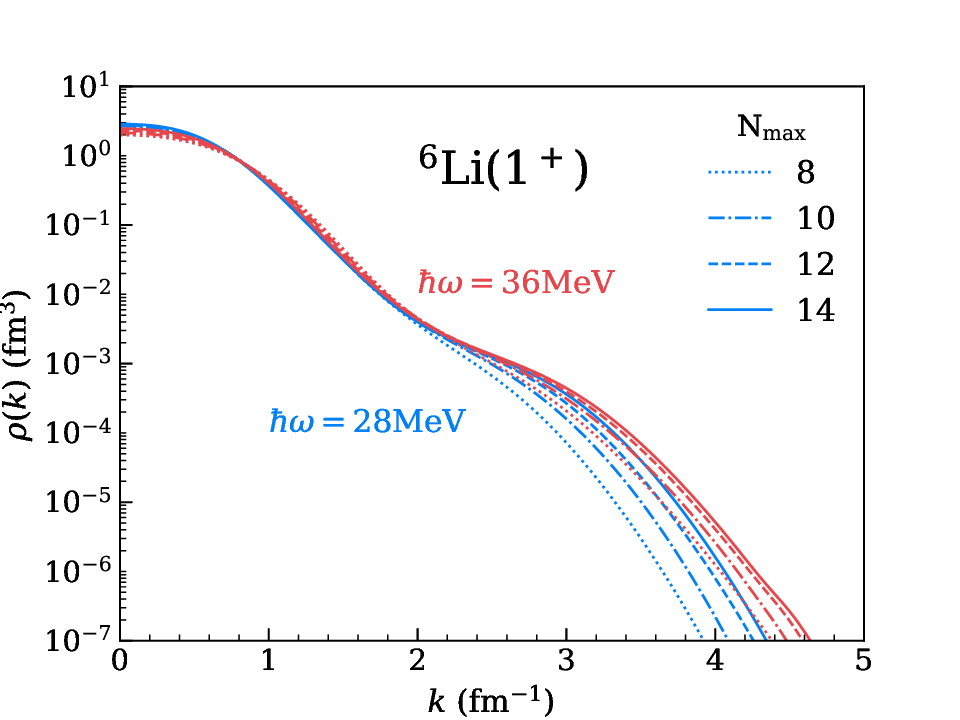}
  \caption{\label{fig:Li6_density}
  The momentum distribution of the  ground state of $^6\mathrm{Li}$, which are obtained by NCSM calculations with the N$^4$LO nonlocal force
   for various $N_{\mathrm{max}}$ at $\hbar \omega$ =28  and 36 MeV, respectively.
  }
\end{figure}

Next we extend the calculations to $^6\mathrm{Li}$,
 for which no results with nonlocal chiral forces have been reported.
In Fig.~\ref{fig:Li6_density} we show the ti densities for two
sets of finite basis spaces, with $\hbar \omega$=28 MeV and 36 MeV, respectively.
The ti densities shown here are extracted directly using Eq.~\eqref{eq:unfoldcal} without the matching procedure as
there are no oscillations at high momentum, in contrast to $^3$He.
Due to the limitation of computational resources, we are only able to reach a truncation at $N_{\mathrm{max}}=14$.
As shown in Fig.~\ref{fig:Li6_density}, the high momentum tail converges more rapidly in a HO basis with $\hbar \omega=36$ MeV than $\hbar \omega =28$ MeV.
Therefore, in the following studies of $A=6$ nuclei, we adopt $\hbar \omega=36$ MeV consistently.
Note that $\hbar \omega=36$ MeV is also the HO basis that minimizes the ground-state energy.

\begin{figure}[t]
  \includegraphics[width=\columnwidth]{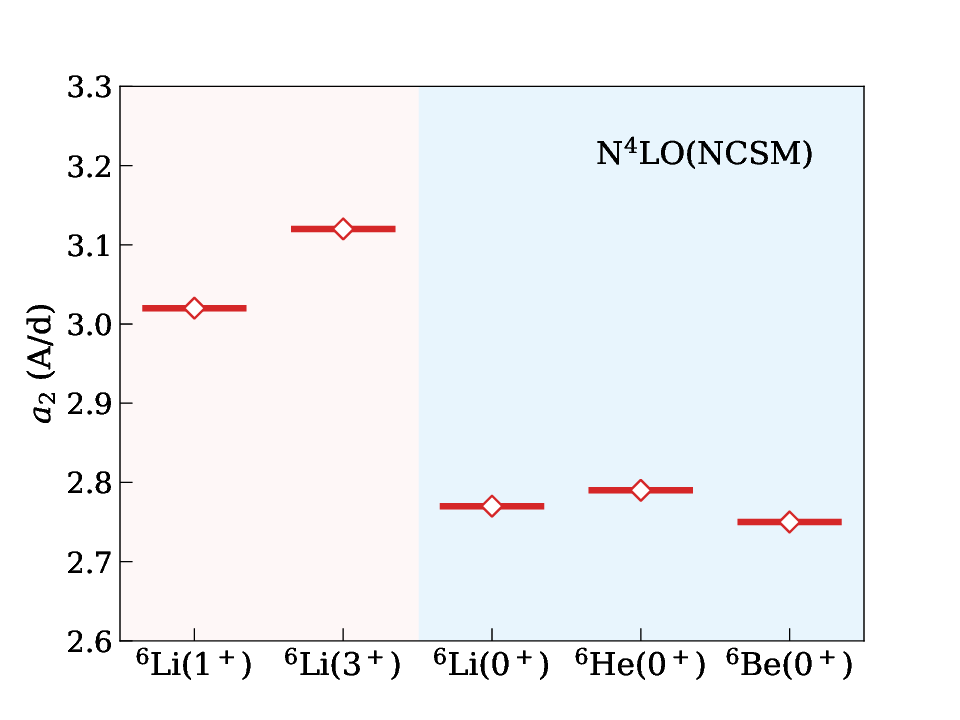}
  \caption{\label{fig:scaling_Li6}
  The scaling factors $a_2$ of different nuclear states in $^6$Li extracted with $k_0=2$ $\mathrm{fm}^{-1}$ using densities from NCSM calculations with $\hbar \omega=36$ MeV.
 Results of 0$^+$ ($T=1$) isobaric triplet states in three neighboring nuclei are also displayed.
  }
\end{figure}

To demonstrate the many-body effects, the scaling factors of  1$^+$ ground state, as well as the subsequent 3$^+$ and 0$^+$ excited states are
shown in Fig.~\ref{fig:scaling_Li6}, in which $\hbar \omega=36$ MeV and $k_0=2$ $\mathrm{fm}^{-1}$ are adopted.
The calculated excited energies of 3$^+$ and 0$^+$ are 2.68 and 4.19 MeV, respectively, which agree reasonably with experimental data 2.186 and 3.563 MeV.
There is some $k_0$ dependence, which is similar to $^3$He,  but the relative order of the scaling factors remains unchanged as $k_0$ varies.
It is shown that $a_2$ becomes larger in 3$^+$ state but decreases significantly at 0$^+$ state.
Usually we expect that less-bound excited states are more spatially spread out and should have reduced SRC.
It has been pointed out that the EMC effect increases with increasing scaled nuclear densities~\cite{cluster}.
Our results
indicate that nuclear states with specific configurations, i.e. different angular momentum and parity, play a significant role
in determining SRC pairs.

To further understand the influences of specific nuclear states,
 we consider the $T=1$  $^{6}\text{He-}^{6}\text{Li-}^{6}\text{Be}$ isobaric triplet 0$^+$ states.
Due to isospin symmetry, these 0$^+$ states are expected to have similar structures, which is
supported by NCSM calculation showing that they exhibit close scaling factors.
This implies that, in these isobaric analog states, certain $np$ pairs in $^6$Li have similar contributions to SRC as
$nn$ pairs in $^{6}\text{He}$ (or $pp$ pairs in $^{6}\text{Be}$), leading to a suppression of SRC pairs.
The scaling factor in  $^{4}$He is calculated to be 3.4 by NCSM.
Then the estimated number of
SRC pairs ($A\times a_2$) in $^{4}$He  is slightly smaller than that of the triplet 0$^+$ states.
Note that for ground states in $^6\mathrm{Li}$ and $^6\mathrm{He}$, calculated $a_2$ with densities from VMC using AV18 agree well with our NCSM results.

\begin{table}[b]
  \caption{
  The extracted  probabilities of $^3S_1$ components  in states of $^{6}\text{Li}$.
  The probabilities of $^3S_1$  is obtained by shell model calculations with  two valence nucleons and  the Moshinsky transformation.
  The configurations of two nucleons and the configuration  components are listed in the 3rd and 4th column, respectively.
  }
  \label{tab:Moshinsky_config}
  \begin{ruledtabular}
  \begin{tabular}{lccc}
      $J^\pi$ &    $^3S_1$ (\%)      &           \text{Configurations}                  &      \text { components }(\%)     \\
      \colrule
      $1^+$            &         36.63           &  $ (\pi 0 p_{3/2})^1 \otimes (\nu 0 p_{3/2})^1$  &               36.82             \\
                       &                         &  $ (\pi 0 p_{1/2})^1 \otimes (\nu 0 p_{3/2})^1$  &               31.57               \\
                       &                         &  $ (\pi 0 p_{3/2})^1 \otimes (\nu 0 p_{1/2})^1$  &               31.57               \\
      $3^+$            &         50.00           &  $ (\pi 0 p_{3/2})^1 \otimes (\nu 0 p_{3/2})^1$  &               100               \\
      $0^+$            &         0.00            &  $ (\pi 0 p_{3/2})^1 \otimes (\nu 0 p_{3/2})^1$  &               76.00              \\
                      &                         &  $ (\pi 0 p_{1/2})^1 \otimes (\nu 0 p_{1/2})^1$  &                24.00              \\
  \end{tabular}
  \end{ruledtabular}
\end{table}

It is striking that 0$^+$ state in $^{6}$Li has similar SRC pairs as $^{6}$He and  $^{6}$Be.
To simplify the analysis of the formation of SRC pairs, we utilize an effective Hamiltonian to describe the $A$=6 system of two valence nucleons around
a $^{4}\mathrm{He}$ core, since the configuration mixing in NCSM framework are too complex.
In this framework, the differences between the states arise solely from the configurations of the two valence nucleons.
The shell model calculations are done within the $\pi(0p_{3/2},0p_{1/2}) \otimes \nu(0p_{3/2},0p_{1/2})$ model space
using the ckpot~\cite{cohen1965effective} effective interaction.
The configuration components are listed in Table~\ref{tab:Moshinsky_config}.
The Moshinsky transformation~\cite{moshinsky1959transformation,buck1996simple} is applied to extract the relative angular momentum  between two valence nucleons
from the single-particle configurations.
The extracted  possibilities of $^3S_{1}$ component are used as an indicator for the abundance of deuteron-like component.
The $^3S_{1}$ components in $1^+$, $3^+$ and $0^+$ states in $^6$Li are 36.63\%, 50\% and 0\%, respectively, which follow the same order as the scaling factors $a_2$.
Therefore we can conclude that the abundance of SRC pairs inside nuclei can be enhanced by the formation of deuteron-like component.
However, $np$ pairs can not form deuteron-like pairs in some specific quantum states.
For example, $^1S_0$ ($T$=1)  rather than $^3S_{1}$ is dominated in these 0$^+$  isobaric analog  states, which has negligible contributions to SRC.
In this case, SRC is not favored although valence protons and neutrons are in the same orbitals.
This reveals that it is insufficient to estimate SRC pairs by only considering the neutron and proton numbers, and their mean-field shell structures.
The isobaric analog states in $^{10}$C-$^{10}$B-$^{10}$Be have similar structures and it is expected that SRC in the ground state 3$^+$ of $^{10}$B would be enhanced.

\emph{Discussions.---}
Nuclear short-range correlations have broad implications in understandings of
 nuclear forces~\cite{schmidt2020probing,korover2014probing}, dense nuclear matter~\cite{frankfurt2008recent,hen2015symmetry,li2018nucleon}, modification of quark-gluon effects and the EMC effect inside nuclei~\cite{emc,clas2019modified,segarra2020neutron}.
In this Letter, nuclear many-body effects in calculations of the scaling factor $a_2$, which characterizes
short-range correlations,
have been demonstrated by {\it ab initio} no-core shell model calculations.
The key step is to unfold the c.m. motion in densities in momentum space.
The high precision N$^4$LO chiral nucleon-nucleon force is adopted and
the results of light nuclei agree reasonably with experiments.
To demonstrate the many-body effects,
the scaling factors of the ground sate and excited sates of $^6$Li are studied,
which shows that the abundance of SRC pairs increases slightly in 3$^+$ state
but decreases significantly in 0$^+$ state, compared to the ground state 1$^+$.
It is striking to find that $a_2$ is close in the 0$^+$ isobaric analog states
in $^6$Li, $^6$Be, and $^6$He, although their neutron and proton numbers are different.
This is explained as that the formation of deuteron-like components is suppressed
in certain nuclear states by analyzing the shell model configurations with the Moshinsky transformation.
Our work highlights that different states of many-body constitutes play a significant role
in the formation of SRC pairs inside nuclei.
This will be useful to guide experiments on SRC in short-lived exotic nuclei or isomeric states,
by using hadronic probes such as the hydrogen target~\cite{proton} and Bremsstrahlung $\gamma$-rays in heavy ion reactions~\cite{xiao}.

\acknowledgments
 We are grateful to useful discussions with F.R. Xu, P. Yin, D. Bai.
 This work was supported by  the
 National Key R$\&$D Program of China (Grant No.2023YFA1606403, 2023YFE0101500),
  the National Natural Science Foundation of China under Grants No.12475118, 12335007.

\bibliographystyle{apsrev4-1}

\begin{thebibliography}{99}

\bibitem{link}
A.W. Denniston, T. Je\v{z}o, A. Kusina, N. Derakhshanian, P. Duwent\"{a}ster, O. Hen, C. Keppel, M. Klasen, K. Kova\v{r}\'{i}k et al.
Phys. Rev. Lett. 133, 152502 (2024).

 \bibitem{hen2017nucleon}
  O. Hen, G. A. Miller, E. Piasetzky, and L. B. Weinstein, Rev. Mod. Phys. 89, 045002 (2017).

  \bibitem{tropiano2021short}
  A. Tropiano, S. Bogner, and R. Furnstahl, Phys. Rev. C 104, 034311 (2021).

  \bibitem{frankfurt1993evidence}
  L. Frankfurt, M. Strikman, D. Day, and M. Sargsyan, Phys. Rev. C 48, 2451 (1993).

  \bibitem{degli2015medium}
  C. C. degli Atti, Phys. Rep. 590, 1 (2015).



  \bibitem{weiss2015generalized}
  R. Weiss, B. Bazak, and N. Barnea, Phys. Rev. C 92, 054311 (2015).

  \bibitem{emc}
  L.B. Weinstein, E. Piasetzky, D. Higinbotham, J. Gomez, O. Hen, and R. Shneor, Phys. Rev. Lett. 106, 052301 (2011).


  \bibitem{clas2019modified}
  B. Schmookler, D. S. Armbruster, T. C. Hsu, and S. P. Roberts, Nature, 566, 354 (2019).

 \bibitem{egiyan2006measurement}
 K. S. Egiyan, N. Dashyan, M. Sargsian, M. Strikman, L. Weinstein, G. Adams, P. Ambrozewicz, M. Anghinolfi, B. Asavapibhop, G. Asryan, and C. Benesh, Phys. Rev. Lett. 96, 082501 (2006).

  \bibitem{fomin2012new}
  N. Fomin, J. Arrington, R. Asaturyan, F. Benmokhtar, W. Boeglin, P. Bosted, A. Bruell, M. Bukhari, M. Christy, E. Chudakov, and J. Watson, Phys. Rev. Lett. 108, 092502 (2012).

  \bibitem{li2022revealing}
  S. Li, R. Cruz-Torres, N. Santiesteban, Z. Ye, D. Abrams, S. Alsalmi, D. Androic, K. Aniol, J. Arrington, and T. Averett, Nature, 609, 41 (2022).

  \bibitem{ryckebusch2019isospin}
  J. Ryckebusch, W. Cosyn, T. Vieijra, and C. Casert, Phys. Rev. C 100, 054620 (2019).

  \bibitem{weiss2018nuclear}
  R. Weiss, R. Cruz-Torres, N. Barnea, E. Piasetzky, and O. Hen, Phys. Lett. B 780, 211 (2018).


  \bibitem{shneor2007investigation}
  R. Shneor, P. Monaghan, R. Subedi, B. Anderson, K. Aniol, J. Annand, J. Arrington, H. Benaoum, F. Benmokhtar, P. Bertin, and C. Carlson, Phys. Rev. Lett. 99, 072501 (2007).

  \bibitem{korover2014probing}
  I. Korover, N. Muangma, O. Hen, R. Shneor, V. Sulkosky, A. Kelleher, S. Gilad, D. Higinbotham, E. Piasetzky, J. Watson, and N. Wiesner, Phys. Rev. Lett. 113, 022501 (2014).

  \bibitem{duer2019direct}
  M. Duer, A. Schmidt, J. Pybus, E. Segarra, A. Hrnjic, A. Denniston, R. Weiss, O. Hen, E. Piasetzky, L. Weinstein, and R. Wiringa, Phys. Rev. Lett. 122, 172502 (2019).

  \bibitem{schmidt2020probing}
  A. Schmidt, J. Pybus, R. Weiss, E. Segarra, A. Hrnjic, A. Denniston, O. Hen, E. Piasetzky, L. Weinstein, N. Barnea, and R. Wiringa, Nature, 578, 540 (2020).

  \bibitem{levinger1951high}
  J. Levinger, Phys. Rev. 84, 43 (1951).

  \bibitem{weiss2015nuclear}
  R. Weiss, B. Bazak, and N. Barnea, Phys. Rev. Lett. 114, 012501 (2015).

  \bibitem{corrfun}
  R. Cruz-Torres, A. Schmidt, G. A. Miller, L. B. Weinstein, N. Barnea, R. Weiss, E. Piasetzky, O. Hen,
  Phys. Lett. B 785, 304(2018).

  \bibitem{cruz2021many}
  R. Cruz-Torres, D. Lonardoni, R. Weiss, M. Piarulli, N. Barnea, D. Higinbotham, E. Piasetzky, A. Schmidt, L. Weinstein, and R. Wiringa, Nat. Phys. 17, 306 (2021).

  \bibitem{wiringa2014nucleon}
  R. B. Wiringa, R. Schiavilla, S. C. Pieper, and J. Carlson, Phys. Rev. C 89, 024305 (2014).

  \bibitem{piarulli2023densities}
  M. Piarulli, S. Pastore, R. Wiringa, S. Brusilow, and R. Lim, Phys. Rev. C 107, 014314 (2023).

  \bibitem{carlson2015quantum}
  J. Carlson, S. Gandolfi, F. Pederiva, S. C. Pieper, R. Schiavilla, K. E. Schmidt, and R. B. Wiringa, Rev. Mod. Phys. 87, 1067 (2015).

  \bibitem{piarulli2016local}
  M. Piarulli, L. Girlanda, R. Schiavilla, A. Kievsky, A. Lovato, L. E. Marcucci, S. C. Pieper, M. Viviani, and R. B. Wiringa, Phys. Rev. C 94, 054007 (2016).

  \bibitem{wiringa1995accurate}
  R. B. Wiringa, V. Stoks, and R. Schiavilla, Phys. Rev. C 51, 38 (1995).

  \bibitem{entem2003accurate}
  D. Entem and R. Machleidt, Phys. Rev. C 68, 041001 (2003).

  \bibitem{entem2017high}
  D. Entem, R. Machleidt, and Y. Nosyk, Phys. Rev. C 96, 024004 (2017).

  \bibitem{ekstrom2013optimized}
  A. Ekstr\"om, G. Baardsen,  C. Forss\'en, G. Hagen, M. Hjorth-Jensen, G. Jansen, R. Machleidt, W. Nazarewicz, T. Papenbrock, J. Sarich et al., Phys. Rev. Lett. 110, 192502 (2013).

  \bibitem{epelbaum2005two}
  E. Epelbaum, W. Gl\"ockle, and U.-G. Mei\ss ner, Nucl. Phys. A747, 362 (2005).

  \bibitem{reinert2018semilocal}
  P. Reinert, H. Krebs, and E. Epelbaum, Eur. Phys. J. A 54, 1 (2018).

  \bibitem{barrett2013ab}
  B. R. Barrett, P. Navr\'atil, and J. P. Vary, Prog. Part. Nucl. Phys. 69, 131 (2013).



  \bibitem{navratil2004translationally}
  P. Navr\'atil, Phys. Rev. C 70, 014317 (2004).


  \bibitem{gloeckner1974spurious}
  D. Gloeckner and R. Lawson, Phys. Lett. B 53, 313 (1974).

  \bibitem{PhysRevC.86.034325}
  C. Cockrell, J. P. Vary, and P. Maris, Phys. Rev. C 86, 034325 (2012).

  \bibitem{giraud2008density}
  B. Giraud, Phys. Rev. C 77, 014311 (2008).


  \bibitem{chen2017short}
J.-W. Chen, W. Detmold, J.~E. Lynn, and A. Schwenk, Phys. Rev. Lett. 119, 262502 (2017).

\bibitem{lynn2020ab}
J. Lynn, D. Lonardoni, J. Carlson, J. Chen, W. Detmold, S. Gandolfi, and A. Schwenk, J Phys. G Nucl. Part. Phys. 47, 045109 (2020).


\bibitem{marcucci2019momentum}
  L. E. Marcucci, F. Sammarruca, M. Viviani, and R. Machleidt, Phys. Rev. C 99, 034003 (2019).

\bibitem{michel2021gamow}
N. Michel and M. P{\l}oszajczak, Gamow Shell Model, 983, Springer (2021).

 \bibitem{marcucci2020hyperspherical}
  L. E. Marcucci, J. Dohet-Eraly, L. Girlanda, A. Gnech, A. Kievsky, and M. Viviani, Front. Phys. 8, 69 (2020).



\bibitem{cluster}
  J. Seely, A. Daniel, D. Gaskell, et al., Phys. Rev. Lett.  103, 202301 (2009).

\bibitem{cohen1965effective}
S. Cohen and D. Kurath, Nucl. Phys. 73, 1 (1965).


\bibitem{moshinsky1959transformation}
M. Moshinsky, Nucl. Phys. 13, 104 (1959).

\bibitem{buck1996simple}
B. Buck and A. Merchant, Nucl. Phys. A600, 387 (1996).

\bibitem{frankfurt2008recent}
L. Frankfurt, M. Sargsian, and M. Strikman,  Int. J. Mod. Phys. A, 23, 2991 (2008).

\bibitem{hen2015symmetry}
O. Hen, B.-A. Li, W.-J. Guo, L. Weinstein, and E. Piasetzky, Phys. Rev. C 91, 025803 (2015).

\bibitem{li2018nucleon}
B.-A. Li, B.-J. Cai, L.-W. Chen, and J. Xu, Prog. Part. Nucl. Phys. 99, 29 (2018).


  \bibitem{segarra2020neutron}
  E. P. Segarra, A. Schmidt, T. Kutz, D. Higinbotham, E. Piasetzky, M. Strikman, L. Weinstein, and O. Hen, Phys. Rev. Lett. 124, 092002 (2020).
\bibitem{proton}
M. Patsyuk,  J. Kahlbow,  G. Laskaris, et al.  Nat. Phys. 17, 693 (2021).


\bibitem{xiao}
Junhuai Xu, Yuhao Qin, Zhi Qin, Dawei Si, Boyuan Zhang, Yijie Wang, Qinglin Niu, Chang Xu, Zhigang Xiao,
Phys. Lett. B 857,139009 (2024).

\end{thebibliography}

\end{document}